\documentclass[a4paper]{jpconf}
\usepackage{graphicx}
\usepackage[export]{adjustbox}
\usepackage{float} %% to put figure where you want rather float around
\usepackage{xurl} %% to break urls into two lines
\usepackage[margin=1in]{geometry}
\usepackage{hyperref}
\hypersetup{
 colorlinks=true,
 citecolor=blue,
 linkcolor=blue,
 filecolor=blue, 
 urlcolor=blue,
 }

\begin{document}

\title{U.S.~CMS - PURSUE (Program for Undergraduate Research SUmmer Experience)}

\author{Tulika Bose$^1$, Sudhir Malik$^2$, Meenakshi Narain$^3$}

\address{$^1$Physics Department, University of Wisconsin Madison, Madison, WI 00682, USA}
\address{$^2$Physics Department, University of Puerto Rico Mayaguez, Mayaguez, PR 00682, USA}
\address{$^3$Department of Physics, Brown University, Providence, RI 02912, USA}

\ead{sudhir.malik@upr.edu}

\renewcommand{\footskip}{10pt} %adjust page number downwards ddm

\begin{abstract}
Students from under-represented populations, including those at minority serving institutions have traditionally faced many barriers that have resulted in their being under-represented in High Energy Physics. These barriers include lack of research infrastructure and opportunities, insufficient mentoring, lack of support networks, and financial hardship, among many others. Recently the U.S.~CMS Collaboration launched a pilot program ``U.S.~CMS - PURSUE (Program for Undergraduate Research SUmmer Experience)'' to address these barriers. A 10-week paid internship program, the very first of its kind in an HEP experiment, was organised during the summer of 2022. Students were selected predominantly from Minority Serving Institutions with no research program in HEP. This pilot program provided a structured hands-on research experience under the mentor-ship of U.S.~CMS scientists from several collaborating institutions. In addition to emphasis on hands-on research, the program offered a set of software training modules for the first few weeks. These were interleaved with a series of lectures every week covering a broad range of topics. The students were exposed to cutting-edge particle physics research and developed a broad set of skills in software, computing, data science, and machine learning. The modality of this program was virtual, due to the unknown circumstances following the pandemic. There is plan to continue the internship program annually, with in-person training and research participation. In this paper, we describe the experience with the pilot program U.S.~CMS - PURSUE.
\end{abstract}

\section{Introduction}

The experiments at the Large Hadron Collider (LHC)\cite{lhc_cern} are one of the big drivers of frontier science and human knowledge. The physics program of the Compact Muon Solenoid (CMS)\cite{cms_experiment} and other LHC based experiments will continue into the 2040s and there are future large HEP experimental endeavors planned on a similar time scale. On a yearly basis, hundreds of next-generation scientists are trained in particle physics by engaging a large number of young researchers (undergraduate, graduate, and postdoctoral researchers) in every aspect of the experiment -- from design, construction to operations, data analysis, and scientific publications. In addition, the recent Snowmass ~\cite{snowmass21} planning process guides planning of the HEP field for a long term future. A common thread that runs across HEP is the development of a future workforce required to carry on the science in the decades to come. This workforce development is essential on all HEP fronts – detector, computing, physics, theory, and accelerator. While we invest in science, it is equally imperative that we integrate in our mission, opportunities for participation and contribution from underrepresented and marginalized populations of our society. Otherwise there exists a danger of being disconnected from such sections of the society, leading to a lack of participation and possible alienation, as well as missed opportunities for scientific discoveries and impact in HEP that have been proven to be enabled by a diverse workforce. In particular, students at Minority Serving Institutions (MSIs)~\cite{hsi_aanapisi} face situations that puts them at a great disadvantage compared to their peers at other institutions, especially when it comes to participation in science, physics and HEP. For example, such students might be as passionate about science as their more advantaged peers, but lack quality academic preparation, need to work to financially support themselves while taking classes, and/or have greater family responsibilities. There is also a lack of academic and career mentorship, and opportunities to look ``unique'' or ``stand out''. Hence they must be offered direct opportunities to engage in Big Science. This pilot program was an attempt to remove such barriers and facilitate building a path for minoritized populations into the STEM workforce, in particular HEP. It aims to address the barriers with a paid and structured research experience complemented with training in software/analysis and a series of lectures covering a broad range of topics. This pilot program named \textbf{U.S.~CMS PURSUE}, is an acronym for \textbf{U.S.} \textbf{CMS} \textbf{P}rogram for \textbf{U}ndergraduate \textbf{R}esearch \textbf{SU}mmer \textbf{E}xperience and it is the very first of its kind in an HEP experiment. This edition of the U.S.~CMS PURSUE program was virtual but future annual iterations are planned to be held in-person. The program will continue to address the lack of diversity in physics/high-energy physics (HEP) due to racial, ethnic or gender identity by focusing on under-represented minority (URM) including Black, Hispanic, and indigenous peoples communities as well as other marginalized communities.

\section{The U.S.~CMS Collaboration}
U.S.~CMS is a collaboration of 54 universities and institutes that work on the CMS Experiment at the LHC. It the largest national group in the CMS Experiment, comprising of 30\% of its collaborators with about 1200 physicists, graduate students, engineers, technicians, and computer scientists. In addition, there are a few hundred undergraduate students who are engaged in its research program. The U.S. collaboration makes significant contributions to nearly every aspect of the CMS detector throughout all phases, including construction, installation, and data-taking. U.S.~CMS also plays a major role in the construction and operation of the experiment's computing facilities and software that is used to analyze the unprecedented amount of data that CMS generates. These highly sophisticated computing tools allow physicists to operate the CMS detector, reconstruct the data, analyze it and, ultimately, make discoveries. 

\subsection{Commitment to Diversity, Equity, and Inclusion} 
The U.S.~CMS Collaboration is committed to empowering its members to develop and achieve their full potential as excellent and innovative particle physicists. The Collaboration acknowledges that diversity, equity, and inclusion (DEI) are fundamental values that impact all aspects of work by the members of our community. Its DEI framework\cite{USCMS-DEI} is based on examining the U.S.~CMS workplace culture and acting to create an inclusive, and collaborative environment that provides a safe and respectful space for diverse voices and perspectives. U.S.~CMS is committed to increasing the representation of women and historically under-represented and marginalized groups in HEP. The U.S.~CMS collaboration has developed broad goals for DEI initiatives, with specific action plans associated with each goal. {\it\textbf{``Identity formation and community engagement"}} is one of the goals of the DEI plan. Providing intentional opportunities to work in diverse groups leads to inclusive research communities and promotes ``physics identity'' formation for colleagues from traditionally underrepresented groups~\cite{WangHazari,TeamUp}. This internship was a step towards realizing this stated goal. Such internships in future would facilitate participation of undergraduate students from non-CMS institutions, specifically by URMs in physics research and encourage U.S.~CMS members to engage with these programs by supervising students, contributing to the planned educational activities, and sharing graduate education opportunities with students.

\subsection{Unique strength and background} The commitment to DEI by U.S.~CMS, complemented by the resources and elements described below, makes it well positioned to continue the U.S.~CMS PURSUE program and its goals. The CMS experiment has not only been at the forefront of HEP with the discovery of the Higgs boson in 2012, but has also been a leader among the LHC experiments to foster a sense of in-reach and outreach for the community. U.S.~CMS collaboration has taken a lead in establishing training programs ~\cite{cmsdas_lpc}~\cite{cmsdas_iop} for jump-starting physics contributions, and has trained thousands of users. These training programs have also been adopted by other CERN experiments. The CMS career programs established to build networking with alumni has led to several pathways for its users to find STEM careers inside or outside academia, and was a prime mover for establishing the CERN wide alumni portal ~\cite{cern_alumni}. It has also encouraged new users to come to HEP by clearing the fog of career opportunities that go far and beyond the specific science that we are engaged with. Many members of the U.S.~CMS collaboration have been have played key roles in establishing a common HEP wide hands-on software training curriculum developed by the HSF (HEP Software Foundation \cite{hsf_training}) and IRIS-HEP ~\cite{irishep_training}. They have attracted students across HEP, Nuclear Physics and Computer Science across the globe to get trained by the HEP community. The curriculum features basics to advance topics that equip students with a broader set of skills that are used in HEP or other areas of research or for getting STEM jobs in industry. Fermilab serves as the hub of U.S.~CMS collaboration and hosts The LHC Physics Center (LPC)~\cite{lpc_fermilab} that was established in 2006 as a regional center of the CMS Collaboration in the U.S. The LPC offers a dynamic and vibrant community of CMS scientists from the U.S. and abroad who play leading roles in analysis of physics data, algorithms and refinement of physics objects, in design, characterisation and construction of the detectors. In addition LPC provides outstanding computing resources and software support personnel. 
%The LPC offers educational workshops in data analysis, advance computing and software, graduate level academic courses related to HEP and organizes conferences and seminar series.

\section{Internship Overview}
\begin{figure}[ht]
\centering
\includegraphics[width=35pc]{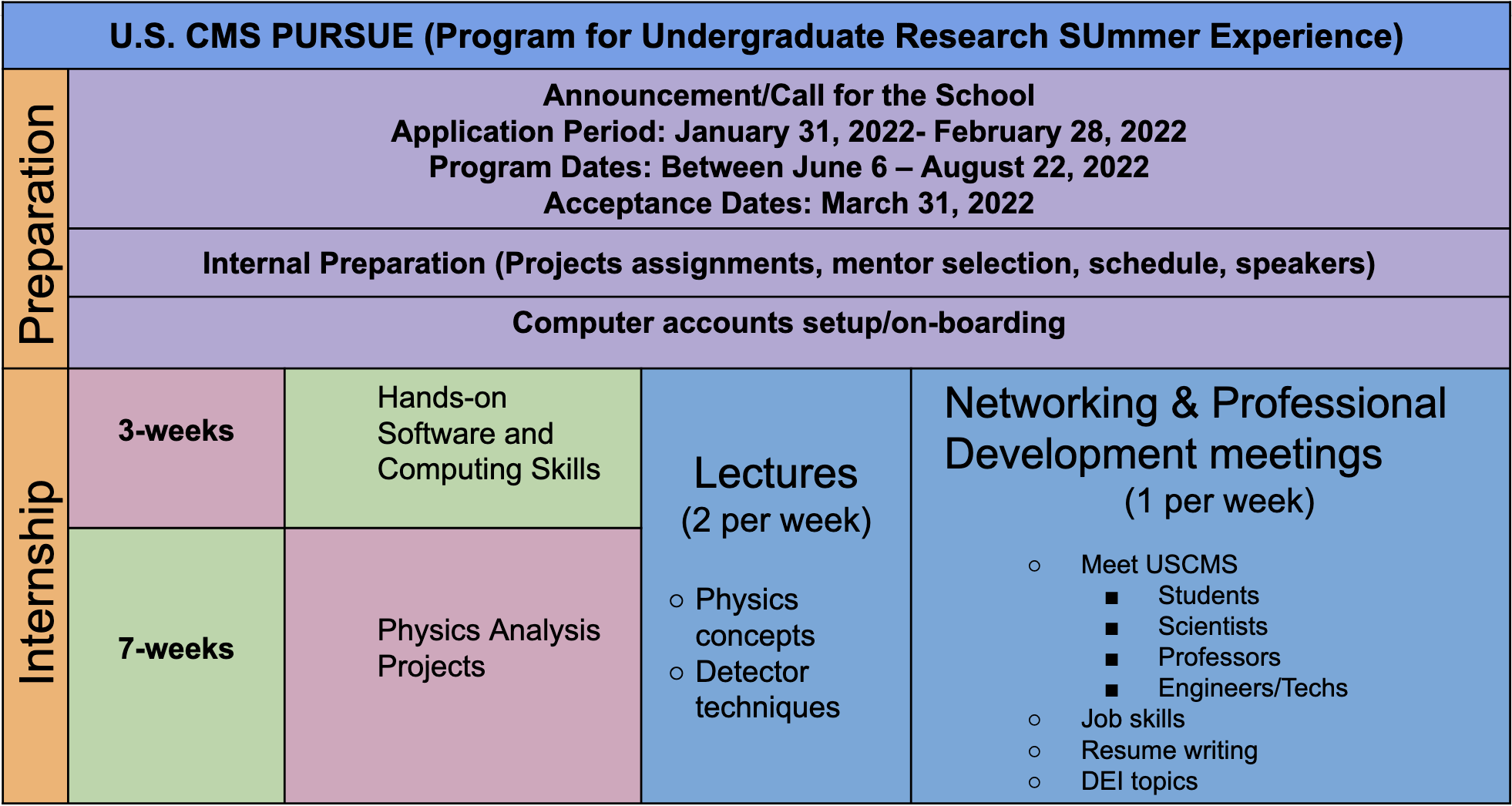}\hspace{5pc}%
\caption{U.S.~CMS PURSUE Program Overview }
\label{fig:program}
\end{figure}

The organization of U.S.~CMS PURSUE internship took about six-months to start after its initial announcement~\cite{uscms_intern_2022}. The application requests were minimal, and included a letter of recommendation, ideally from a STEM faculty member at the current institution, an unofficial transcript including course titles and overall GPA, a one-page resume and an essay describing interest in the internship program. The essay contained applicant's background, skills, strengths, scientific topics that appeal to them, previous research experience, if any; what they may want to pursue as a future career; and what benefits they would like to gain from this program. Over 100 applications were received and a selection rubric was created and fine-tuned to reduce the risk of implicit bias. Finally, 16 participants were selected and matched with a Physics Project. In parallel and leading up to the start of the program, internal preparations were made with respect to projects assignments of interns. 
%Program facilitators selected the instructors, mentors and lecturers, and organized the weekly schedule of internship. 
Facilitators were responsible for the overall organisation of the program. This required defining the program, making a call to intern applications, reviewing and selecting interns from application pool, detailed timetable of the program and work plan of interns and recruitment of mentors, instructors and lecturers. Instructors were scientists and primary academic drivers who were responsible for the software training and physics projects. The ratio of instructor to interns was 1:1. Mentors helped interns throughout if they had any cluster login, connection, operating system or software programming issues. Lecturers were senior graduate students, postdocs, scientists/professors who gave lectures on a variety of topics that complemented the program goals. An agenda was created to reflect the daily time table and overall schedule of the interns. Given the virtual nature of the internship a communication portal was setup using Slack. The Slack workspace enabled students to effectively communicate with the entire team in case of any software, physics project or general questions. Mentors monitored the Slack channel to watch for any issues or questions posted by the interns. General communication was done by a group email setup by the facilitators where everyone subscribed. An overview of program is shown in Figure~\ref{fig:program}.

\section{The Program} The curriculum of the 10-week internship consisted of components as described below, with a focus to provide the interns an experience and exposure to the world of scientific discovery. While the physics project work introduced hands-on experience with physics analysis, the hands-on software skills training was designed to help with performing the physics projects and are applicable STEM-wide in academic or a non-academic career paths. Hands-on experience was interleaved with 3 lectures per week on topics ranging from basics of particle physics, to Discovery Physics, handling Big Computing/Software platforms, and introduction to the basics of a particle physics detectors, including discovery capabilities of the state-of-the sub-detectors of CMS apparatus. Every week the internship students were given the opportunity to network with peers, meet with U.S.~CMS scientists (graduate students, postdoctoral fellows, research scientists and faculty). During the internship, a few workshops on professional development opportunities, and topics relating to diversity and mentorship in particle physics were conducted. As the internship was entirely remote and conducted via zoom, the lecture sessions and usage of Slack helped interns connect and learn from each other. The usage of the Slack portal helped students access support throughout the internship from their research and program mentors, whether it was to seek solutions to questions about their research or basic computing and physics, or questions on future career opportunities or to get acquainted with their peers and share experiences with the program.

 \begin{itemize}
 \item \textbf{Software training:} First three-weeks of the internship were devoted to a software curriculum on basic computational \& software tools and data-science methods required for the physics projects. Software is a fundamental tool to any scientific work and certainly in HEP where it is used for data processing, event reconstruction, simulations and physics data analysis among other tasks. The software curriculum comprised of learning Linux OS skills, basics of working on a computing cluster, Python programming, C++ programming, ROOT framework, Machine Learning and collaborative code development and version control via Github. 
 
 \item \textbf{Physics Projects:} The Physics projects were matched by the broad interests expressed by the interns during the application process and varied from searches on supersymmetry, dark matter, exotic particles to particle reconstruction, detector performance identifying physics objects like taus, muons and electrons, usage of GPUs, Machine Learning and data management tools. The projects exposed students to usage of software, computing and grid tools to analyze data from the CMS experiment. It enabled an understanding of the workings of the nature at its most fundamental levels - interactions of fundamental particles and fundamental forces.
 
 \item \textbf{Lecture series:} About 2 lectures per week formed part of the curriculum on carefully chosen topics - Physics searches at CMS, CMS sub-detectors: \textit{Tracker, Calorimeter, Muon} etc., meeting a U.S.~CMS grad student/postdoc/scientist, importance and strength of DEI, networking and resume building for career in STEM or industry. While physics and detector lectures helped interns relate to their projects and importance of science, meetings with U.S.~CMS grad students, postdocs and scientists enabled them to familiarize with various career paths and progression. Additionally, they were exposed to the challenges, passion, motivation and dedication required to do science by meeting (online) and interacting with CMS scientists. Networking, skill building and resume tips sessions emphasized the importance to develop and improve skill set, stay abreast with trends in job opportunities in academia or STEM industry, meet prospective mentors and partners, and how to gain access to the necessary resources that foster career development. Q\&A during lectures was strongly encouraged and almost 1/3 of the time was reserved for discussions during the lecture sessions.
 
 \item \textbf{Evaluation:} 
 During the initial weeks of the internship, interns were requested to meet with their research mentors, discuss their projects, and then present their understanding of their projects in an exercise called \textit{Five-slides in five-minutes}. This presentation exercise helped the research mentors to evaluate initial level and grasp of potential physics work by the student, and in parallel gave the students the space to get familiar with ideas and skills they would learn during the internship and instilled the confidence to present their initial understanding before the experts. At the end of the internship students were required to make a poster based on their physics research projects and also give oral presentations on the poster. Each intern's final presentation was reviewed for feedback by their peers, mentors and reviewers. In addition, mentors provided a detailed review of the entire duration of work of their respective mentees (interns).
 
\section{Feedback} Feedback of the internship program was collected in two stages. An initial feedback survey was provided to the interns three weeks into the program. It had three goals - (1) evaluation of the program thus far; (2) level of understanding of their physics project by the interns; and (3) re-calibrate or adjust the expectations by the research mentors for remaining duration, if needed. By end of the three week period, most software training lessons on Linux, Python, Github and C++ were complete, a dozen lectures on physics, CMS (sub)detectors, DEI/networking were already given and regular project discussions with respective mentors had started to take shape. There were also one-on-one conversations between the research mentors and facilitators with interns during this three week period. In addition the "Five-slides in five-minutes" presentation on the abstracts of their projects helped the research mentors gain insight on the initial level of understanding of physics and skills of their mentees.

Some results from this initial feedback (after three-weeks of internship) are shown below in Figures~\ref{fig:learning} \ref{fig:resource} \ref{fig:mentor}. In the feedback comments one intern remarked - \textit{I'm learning a lot and having a lot of fun. This is an experience I value a lot and won't forget for a long time - it's probably something I'm going to look back upon as a career-shaping experience in the future} and another responded on what they did they not understand well enough thus far -\textit{It's the mechanisms of how some computer based things work. Like, I'm working on data conversion and I do not know how exactly the conversion takes place, I'm executing a file but I want to be able to understand what it does too (No complaints on the general program, this is just something I'm going to work on by myself)}. Some of the useful information and observation extracted is as follows:

\begin{figure}[ht]
\centering
\includegraphics[width=35pc, frame]{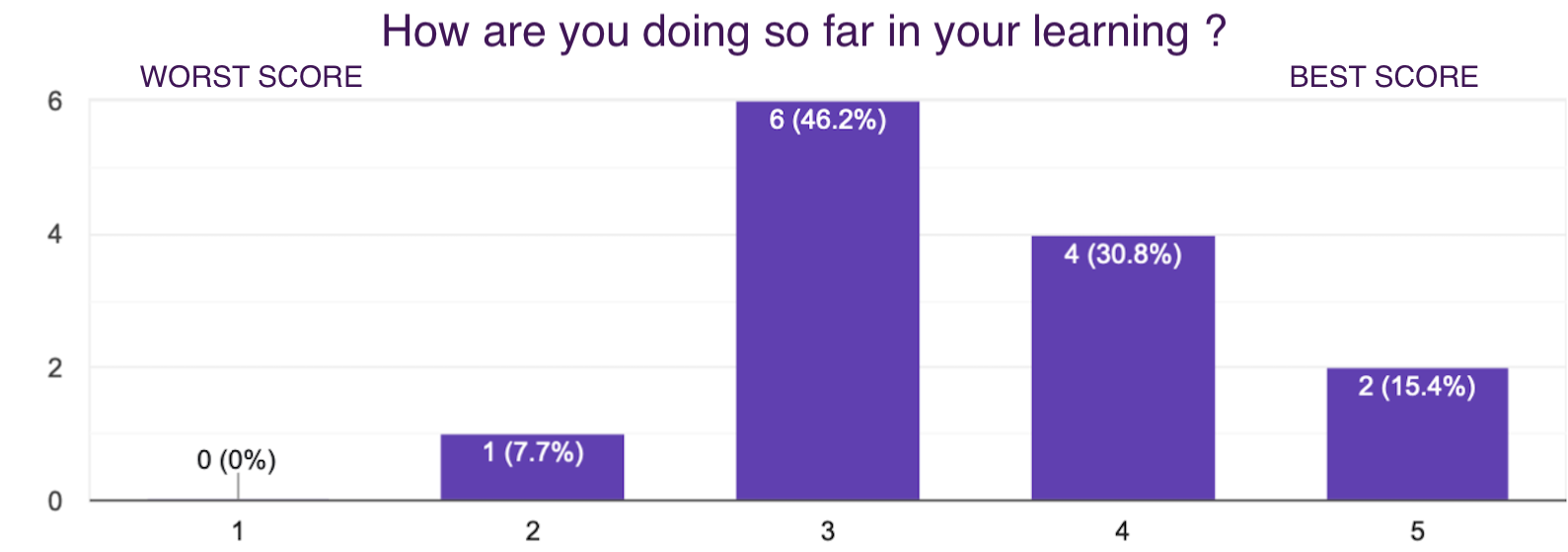}\hspace{5pc}%
\caption{1 is worst and 5 is the best}
\label{fig:learning}
\end{figure}

\begin{figure}[ht]
\centering
\includegraphics[width=35pc, frame]{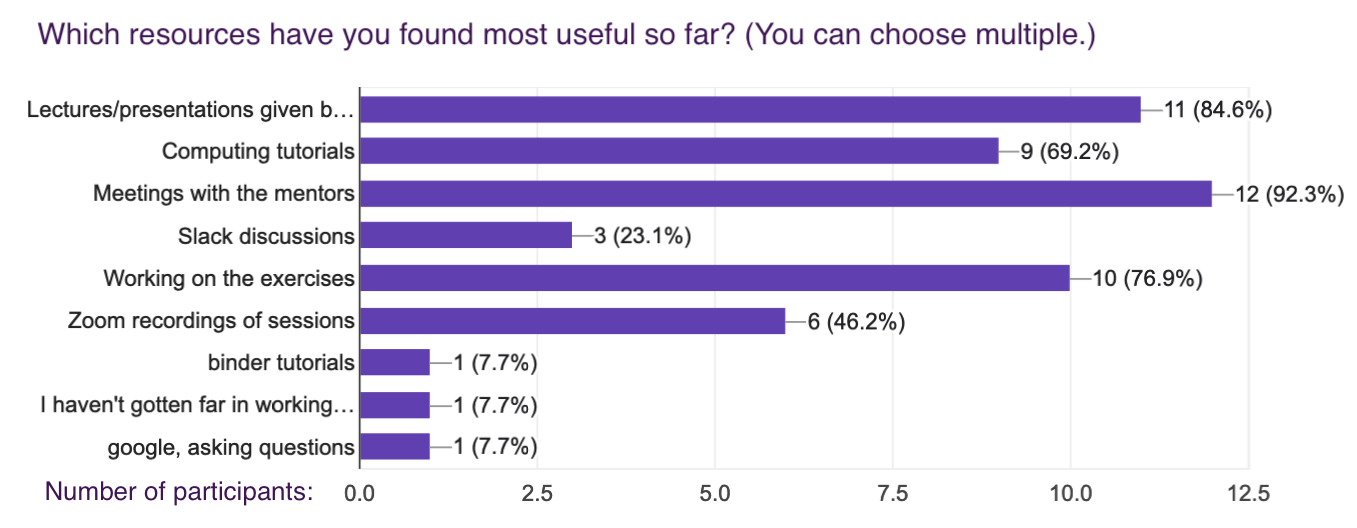}\hspace{5pc}%
\caption{ horizontal axis is no. of interns }
\label{fig:resource}
\end{figure}

\begin{figure}[ht]
\centering
\includegraphics[width=35pc, frame]{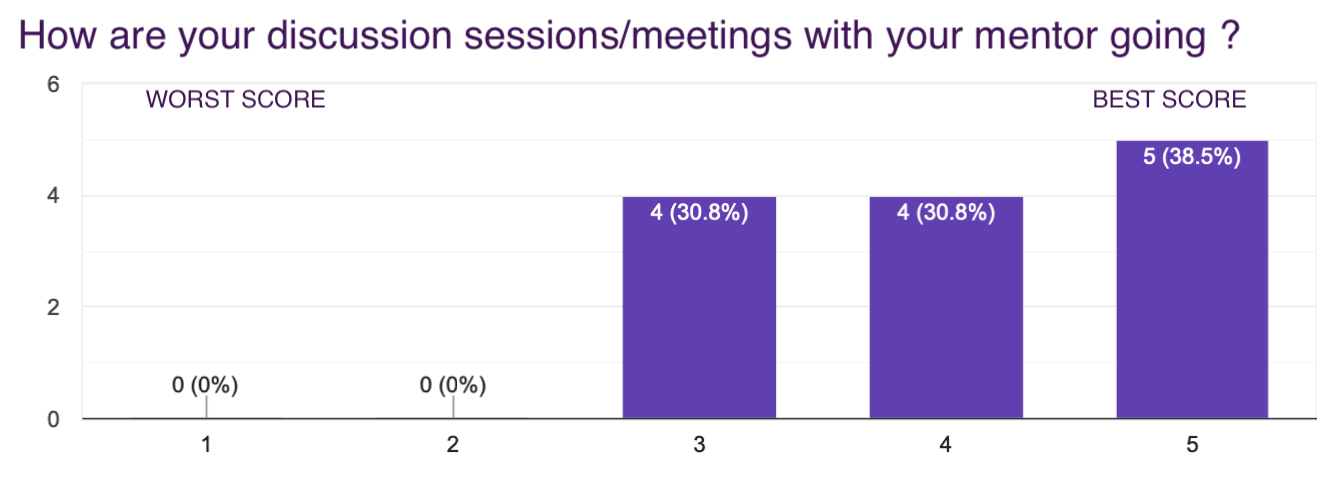}\hspace{5pc}%
\caption{1 is worst and 5 is the best}
\label{fig:mentor}
\end{figure}
 
 \begin{itemize}
 
 \item Internship program is critical to give opportunities to students especially at MSIs and PUIs and increase the STEM workforce.
 \item It is absolutely essential to have a strong in-person component of the program to build a sense of community. One intern remarked \textit{I would like to interact more with the other mentees outside of talks/training} and a second said \textit{more social opportunities between interns} and a third one said \textit{I personally don’t do well with learning virtually. I’m a hands on individual, so between now and the end of the internship, I hope to improve on this.}.
 \item Time zone-differences between interns and a very few Europe based U.S.~CMS mentors limited the zoom time availability.
 \item Most interns were good learners but clearly lacked such opportunities and challenges at their institutions. Peer effect and collective sessions (though zoom limited) did provide mutual motivation. 
 \item Some interns did not have access to personal laptops and they were promptly shipped one before the onset of the program
 \item Exposure to software skills prior to the internship was none to negligible and hence an initial phase of software training is required to complete projects with understanding and success.
 \item Most interns were not appreciative of software skills initially. One intern remarked \textit{I do believe we are not getting a lot of time to actually do research. It seems like a lot of learning and rather than research.} but quickly realised its indispensability for their projects. Hence it is important to emphasize in the initial call for the internship the software skills to be taught for their project's successful completion, Software skills are virtually non-existent from undergrad curriculum even at research focused universities.
 \item Interns need a more guided approach in terms of weekly tasks and expectations to feel incremental achievements.
 \item All interns found their mentors as exceptional and patient teachers
 \item "Five-slides in five-minutes" presentation early on to see initial grasp on the project was useful and one intern remarked \textit{Putting the presentation together was very useful as it made me work through my current understanding of my project and put it together in a coherent narrative}.
 
 \end{itemize}
 \end{itemize}

\begin{figure}[ht]
\centering
\includegraphics[width=35pc, frame]{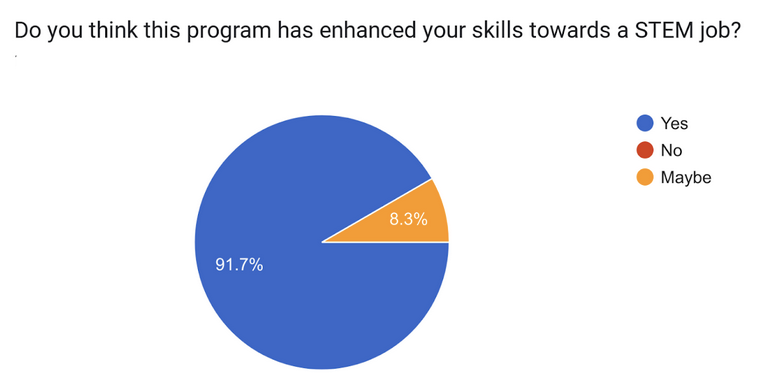}\hspace{5pc}%
\caption{ }
\label{fig:skills}
\end{figure}

\begin{figure}[ht]
\centering
\includegraphics[width=35pc, frame]{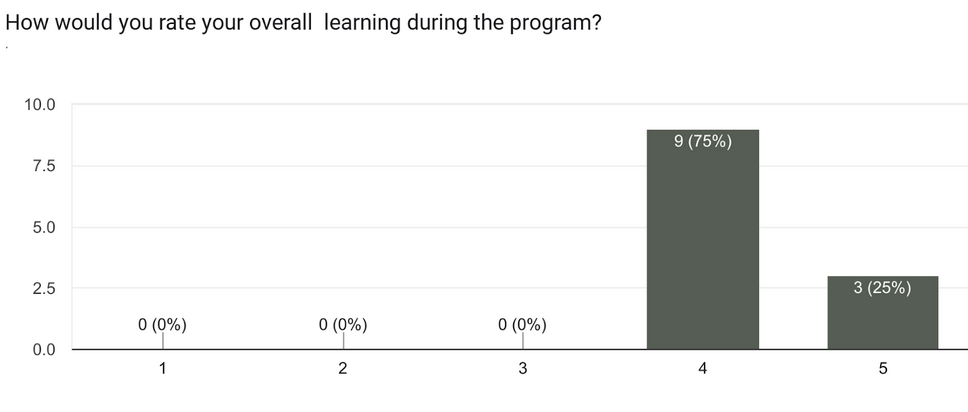}\hspace{5pc}%
\caption{1 is worst and 5 is the best}
\label{fig:rating}
\end{figure}

\begin{figure}[ht]
\centering
\includegraphics[width=35pc, frame]{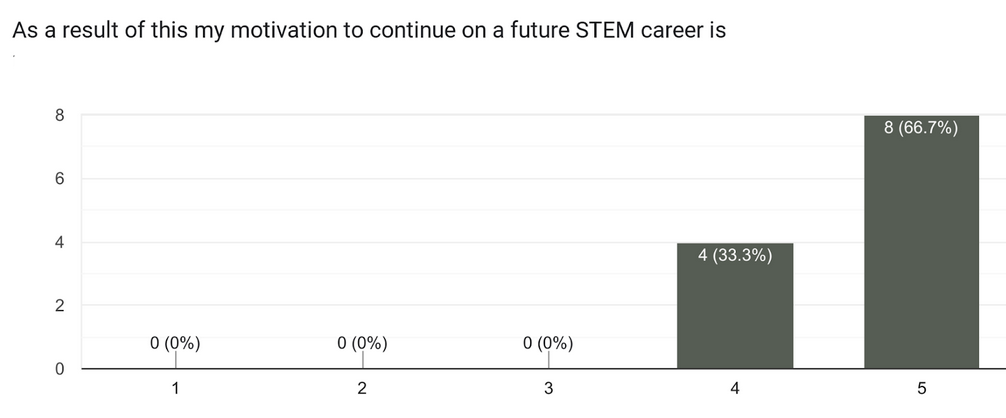}\hspace{5pc}%
\caption{1 is worst and 5 is the best}
\label{fig:motivation}
\end{figure}

Selected feedback from post internship survey is shown in Figures~\ref{fig:skills} \ref{fig:rating} \ref{fig:motivation}. The value served by the U.S.~CMS PURSUE internship is clearly epitomised in ~\cite{ralphortiz} and these figures, including the quality of the internship offered by this pilot program

\section{Conclusions}
The U.S.~CMS PURSUE internship provides an opportunity for the U.S.CMS Collaboration to implement its goal of ``Identity formation and community engagement'' within the DEI framework. It demonstrates that efforts providing opportunities for research and engaging them in research enables mitigating the barriers faced by the under-represented populations and students from minority serving institutions. The future of STEM fields and HEP in particularly is entwined with such continuous efforts. The U.S.~CMS PURSUE internship is a unique hands-on learning experience about the intricate and complex structure of the discovery science at the fore-frontier of human knowledge offered by the CMS Experiment. The interns learnt about HEP, the scientific techniques and computing software that have STEM-wide applications, beyond HEP. The feedback from interns provides a strong incentive and rationale to develop this program further. 
%The detailed feedback about parts of the program is very valuable to re-calibrate the future editions the internship.
We are grateful to U.S.~CMS Operations Program for providing funding for this internship. 

\section*{Acknowledgements}
The authors thank all U.S.~CMS colleagues for their support in mounting a successful program. In particular, we thank the following participants who joined the effort in introducing HEP to a wider audience and take an important step towards trying to enhance the diversity of the HEP community:
Lothar Bauerdick, Ken Bloom, Guillermo~Fidalgo ~Rodriguez, Darin Acosta, Todd Adams, Douglas Berry, Agni Bethani, Kevin Black, Johan Bonilla, Bruno Coimbra, Susan Dittmer, Jay Dittmann, Javier Duarte, Pieter Everaerts, Patrick Gartung, Sergei Gleyzer, Ulrich Heintz, Julie Hogan, Bo Jayatilaka, Andy Jung, Keti Kaadze, Georgios Krintiras, Charis Koraka, Amit Lath, Don Lincoln, Daniel Li, Carl Lundstedt,  Jingyu Luo, Devin Mahon, Marco Mambelli, Guenakh Mitselmakher, Abdollah Mohammadi,  Scarlet Nordberg, Isabel Ojalvo, Christopher Palmer, Alexx Perloff, Joseph Reichert, Oksana Shadura, Farrah Simpson, Indara Suarez, Marguerite Tonjes, Benjamin Tovar, Jieun Yoo, Fengwangdong Zhang. We also wish to thank and acknowledge support of U.S. CMS Operations grant NSF-2121686, “U.S. CMS Operations at the Large Hadron Collider”.

\section*{References}

\typeout{}
\bibliography{iopart-num}

\providecommand{\newblock}{}
\begin{thebibliography}{10}
\expandafter\ifx\csname url\endcsname\relax
  \def\url#1{{\tt #1}}\fi
\expandafter\ifx\csname urlprefix\endcsname\relax\def\urlprefix{URL }\fi
\providecommand{\eprint}[2][]{\url{#2}}

\bibitem{lhc_cern}
{The Large Hadron Collider}
  \url{https://home.cern/science/accelerators/large-hadron-collider}
  \urlprefix\url{https://home.cern/science/accelerators/large-hadron-collider}

\bibitem{cms_experiment}
{CMS Experiment} \url{https://home.cern/science/experiments/cms}
  \urlprefix\url{https://home.cern/science/experiments/cms}

\bibitem{snowmass21}
{Welcome to Snowmass} \url{https://snowmass21.org/}

\bibitem{hsi_aanapisi}
 2022 Minority serving institutions (msi) list
  \url{https://orise.orau.gov/msipp/documents/approved-msi-school-list.pdf}
  \urlprefix\url{https://orise.orau.gov/msipp/documents/approved-msi-school-list.pdf}

\bibitem{USCMS-DEI}
{U.S. CMS Diversity, Equity and Inclusion Plan}
  \url{https://drive.google.com/file/d/1cwtXUL2FS4momxXYHsPbvMYpbKUqrBEj/view}

\bibitem{WangHazari}
Wang J and Hazari Z 2018 {\em Phys. Rev. Phys. Educ. Res.\/} {\bf 14}(2) 020111
  \urlprefix\url{https://link.aps.org/doi/10.1103/PhysRevPhysEducRes.14.020111}

\bibitem{TeamUp}
{The American Institute of Physics National Task Force to Elevate African
  American Representation in Undergraduate Physics \& Astronomy (TEAM-UP)
  Report.}
  \url{https://www.aip.org/sites/default/files/aipcorp/files/teamup-full-report.pdf}

\bibitem{cmsdas_lpc}
{CMS Data Analysis School (CMSDAS), CMS Physics Object School (CMSPOS) and CMS
  Upgrade School (CUPS)}
  \url{https://lpc.fnal.gov/programs/schools-workshops/cmsdas.shtml}
  \urlprefix\url{https://lpc.fnal.gov/programs/schools-workshops/cmsdas.shtml}

\bibitem{cmsdas_iop}
{N De Filippis, L Bauerdick, J Chen, E Gallo, B Klima, , S Malik, M Mulders, F
  Palla, G Rolandi} 2017 {\em Journal of Physics: Conf. Series\/} {\bf 898} pp
  26--27
  \urlprefix\url{https://iopscience.iop.org/article/10.1088/1742-6596/898/10/102015/pdf}

\bibitem{cern_alumni}
{CERN Alumni Portal} \url{https://alumni.cern/}
  \urlprefix\url{https://alumni.cern/}

\bibitem{hsf_training}
{HSF} training
  \url{https://hepsoftwarefoundation.org/workinggroups/training.html}
  \urlprefix\url{https://hepsoftwarefoundation.org/workinggroups/training.html}

\bibitem{irishep_training}
{IRIS-HEP}: Training, education and outreach
  \url{https://iris-hep.org/ssc.html}
  \urlprefix\url{https://iris-hep.org/ssc.html}

\bibitem{lpc_fermilab}
{LHC} physics center at fermilab \url{https://lpc.fnal.gov/}
  \urlprefix\url{https://lpc.fnal.gov/}

\bibitem{uscms_intern_2022}
{US CMS Undergraduate Internship}
  \url{https://internships.fnal.gov/u-s-cms-undergraduate-internship/}

\bibitem{ralphortiz}
{Ralph Ortiz III - USCMS PURSUE experience }
  \url{https://www.linkedin.com/posts/rafaelortiziii_cern-fermilab-cms-activity-6967543513278287872-IRpF?utm_source=share&utm_medium=member_desktop}

\end{thebibliography}

%%\section*{References}
%%\begin{thebibliography}{9}
%%\bibitem{iopartnum} IOP Publishing is to grateful Mark A Caprio, %%Center for Theoretical Physics, Yale University, for permission to %%include the {\tt iopart-num} \BibTeX package (version 2.0, December %%21, 2006) with this documentation. Updates and new releases of {\tt %%iopart-num} can be found on \verb"www.ctan.org" (CTAN). 
%%\end{thebibliography}

\end{document}